\documentclass[conference]{IEEEtran}
\IEEEoverridecommandlockouts

\usepackage{cite}
\usepackage{amsmath,amssymb,amsfonts}
\usepackage{algorithmic}
\usepackage{graphicx}
\usepackage{multirow}
\usepackage[utf8]{inputenc}
\usepackage{textcomp}
\usepackage{hhline}
\usepackage{booktabs}
\usepackage{colortbl}
\usepackage{xcolor}
\usepackage{hyperref}
\usepackage{amsmath}

\def\BibTeX{{\rm B\kern-.05em{\sc i\kern-.025em b}\kern-.08em
    T\kern-.1667em\lower.7ex\hbox{E}\kern-.125emX}}
\usepackage{cite}
\def\BibTeX{{\rm B\kern-.05em{\sc i\kern-.025em b}\kern-.08em
    T\kern-.1667em\lower.7ex\hbox{E}\kern-.125emX}}

\begin{document}
\title{Audio-Visual Embedding for Cross-Modal Music Video Retrieval through Supervised Deep CCA}
\author{
\IEEEauthorblockN{Donghuo Zeng}
\IEEEauthorblockA{\textit{National Institute of Informatics} \\
Tokyo, Japan, SOKENDAI \\
zengdonghuo@nii.ac.jp
}

\and
\IEEEauthorblockN{Yi Yu}
\IEEEauthorblockA{\textit{National Institute of Informatics} \\
Tokyo, Japan, SOKENDAI \\
yiyu@nii.ac.jp}

\and
\IEEEauthorblockN{Keizo Oyama}
\IEEEauthorblockA{\textit{National Institute of Informatics} \\
Tokyo, Japan, SOKENDAI \\
oyama@nii.ac.jp
}
}
\maketitle

\begin{abstract}
Deep learning has successfully shown excellent performance in
learning joint representations between different data modalities. Unfortunately, little research focuses on cross-modal correlation learning where temporal structures of different data modalities, such as audio and video, should be taken into account. Music video retrieval by a given musical audio is a natural way to search and interact with music contents. In this work, we study cross-modal music video retrieval in terms of emotion similarity. Particularly, an audio of an arbitrary length is used to retrieve a longer or full-length music video. To this end, we propose a novel audio-visual embedding algorithm by Supervised Deep Canonical Correlation Analysis (S-DCCA) that projects audio and video into a shared space to bridge the semantic gap between audio and video. This also preserves the similarity among audio and visual contents from different videos with the same class label and the temporal structure. The contribution of our approach is mainly manifested in the two aspects: i) We propose to select top k audio chunks by attention-based Long Short-Term Memory (LSTM) model, which can represent good audio summarization with local properties. ii) We propose an end-to-end deep model for cross-modal audio-visual learning where S-DCCA is trained to learn the semantic correlation between audio and visual modalities. 
Due to the lack of music video dataset, we construct 10K music video dataset from YouTube 8M dataset.  
Some promising results such as MAP and precision-recall show that our proposed model can be applied to music video retrieval.
\end{abstract}


\section{INTRODUCTION}
\begin{figure}
\centering
\includegraphics[width=0.48\textwidth]{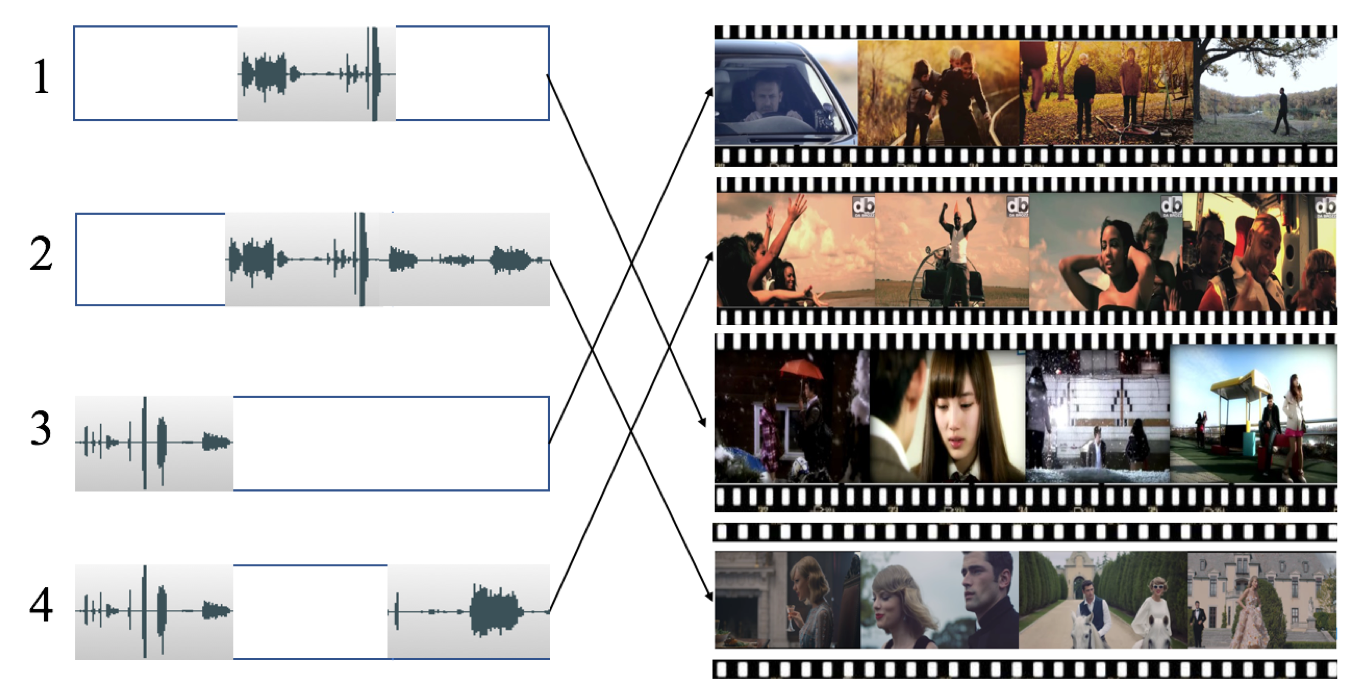}
\caption{\label{figure-1} Overview of music video retrieval: Select one or more representative audio chunks as query to find similar music video, based on content similarity.}
\label{figure1}
\end{figure}

Deep cross-modal learning is a very important research topic in the area of multimedia and computer vision, with the goal of learning joint representation between different data modalities such as image-text~\cite{wang2016learning, yu2018category} and audio-lyrics~\cite{yu2017deep}. In the cross-modal music video retrieval, taking a piece of music audio segment to retrieve visual contents is a natural way to find an interesting music video that facilitates and improves people's music experiences. Let us imagine the scenario: when  a user sits in the bar, a song attracts his attention. He instantly records the song by his cellphone and with this as query finds semantically similar music videos, as shown in Fig.~\ref{figure1}. Correlation learning between visual and audio sequences is non-trivial. However, little work has contributed to this task where temporal structures of different modalities should be considered.

The large volumes of music videos emerged in the Internet provide a nice opportunity for us to learn the correlation between visual and audio temporal sequences. A music video contains visual and audio modalities, which are embedded in musical temporal sequences to express music theme and story. Moreover, as a special form of expression, a music video also conveys strong feelings and emotions, which are semantically contained in audio and visual modalities. That is to say, music emotion is delivered by both audio and visual modalities in music video. This motivates us to learn a joint embedding space where music audio and visual contents are assumed with same semantically meaning.

In this work, we study how to use audio to retrieve music video under a realistic situation: with a segment of music audio that has a variable length as a query, the system automatically finds the music video that is similar to this audio with regard to emotions. In other words, an audio with an arbitrary length can retrieve a longer or full-length music video. It is natural for users to search music video in this way. However, this is a challenging research issue because audio and video are different modalities that have different low-level features with different properties of temporal structures. To this end, we propose a novel audio-visual embedding algorithm by Supervised Deep Canonical Correlation Analysis (S-DCCA) that projects audio and video into a joint feature space to bridge the gap across different modalities. This also preserves the similarity among audio and visual contents from different videos with the same class label and the temporal structure. In addition to selecting 10K music video data from the YouTube-8M dataset, most importantly, several contributions are made in this paper as follows:

i) To the best of our knowledge, this is the first work that studies how to retrieve a full video by an audio having a variable length.

ii) We propose to select k representative audio chunks based on emotion features extracted by a Long Short-Term Memory (LSTM)-based attention model, which serve as audio summary meanwhile conserving the temporal structure.

iii) We propose an end-to-end deep architecture for cross-modal audio-visual embedding where S-DCCA is trained to learn the semantic correlation between audio and visual modalities.

iv) Evaluation demonstrates that our deep model has competitive performance compared with state-of-the-art approaches.

The rest of this paper is structured as follows. Section II introduces related work on deep cross-modal embedding and multimedia retrieval. Section III presents the architecture of our model and Section IV reports the experimental results. Finally, Section VI draws the conclusion and points out future work.

\section{RELATED WORK}
Cross-modal music retrieval intensively focuses on studying music and visual modalities\cite{yu2012automatic, acar2014understanding, mayer2011analysing, brochu2003sound, shah2014advisor, gillet2007correlation}. Similarity between audio features extracted from songs and image features extracted from the album covers are trained by a Java SOMToolbox framework in \cite{mayer2011analysing}. Based on this similarity, people can easily manage a music collection and utilize album cover as visual content to search a music
song from music dataset. Based on multimodal mixture models, a statistical method is applied to jointly model music, images, and text \cite{brochu2003sound} for facilitating music multimodal retrieval. The sensor data streams are mapped to a geo-feature. A visual feature is calculated from video content. With the trained $SVM^{hmm}$ model, mood tags associated with visual-aware likelihood are generated. Then, the likelihoods of the mood tags associated with location information and video content are combined by late fusion. Mood tags with large likelihoods are regarded as scene moods of this video. Finally, the songs matching user's listening history are extracted as personalized recommendations \cite{shah2014advisor}. To learn the semantic correlation between music and video, a method to choosing features and statistical novelty based on kernel methods \cite{gillet2007correlation} is suggested to segment music song. Co-occurring changes from audio and video in music videos can be found, where the correlations can be applied to cross-modal audio-visual music retrieval. 

The key idea of cross-modal correlation learning is to learn a joint space where different modalities can be correlated semantically. In particular, recent progresses mainly focus on cross-modal learning between text and image such as \cite{jiang2016deep, yu2016video}. Most existing deep architectures with two sub-networks exploit pre-trained convolutional neural network (CNN) \cite{simonyan2014very} as image branch \cite{yu2017venuenet} and utilize pre-trained text-level embedding model \cite{lau2016empirical} or hand-crafted feature extraction such as bag of words \cite{jiang2016deep} as text branch. Then image and text features are projected to the shared space to compute a ranking loss function by a feed-forward way. Image-text benchmarks such as \cite{lin2014microsoft, rasiwasia2010new} are used to evaluate the performances of cross-modal matching and retrieval.

Existing deep cross-modal retrieval methods have two properties: i) little work related to cross-modal correlation learning takes into account temporal structure of different modal data. ii) Pre-trained models are directly used to extract image or text features. Distinguished from existing deep cross-modal retrieval architectures, this work takes into account temporal structures to learn the correlation between audio and video for enabling  cross-modal music video retrieval, where sequential audio and visual contents are projected to the same canonical space. An end-to-end neural network architecture with two-branch sequential structures for audio and video is investigated. Most importantly, we propose a novel method that extracts representative chunks from audio, which is able to summarize audios with different lengths. In addition, we propose a supervised deep CCA method to learn their semantic correlation.

\section{ARCHITECTURE}
Ideally, continuous audio segments (called chunks in this paper) , which are short enough, have the same music property, such as emotion attribute. This motivates us to equally divide a long audio sequence into chunks with the same length. Then, the emotion information of each chunk is computed, and the best chunks with the most attention intensity are used to represent the whole audio sequence. By the cross-modal correlation between the best audio chunks and visual features, the most similar videos can be found.

\subsection{Neural Attention Modeling}
The main part of attention computation is realized by the Long Short Term Memory (LSTM) networks model ~\cite{hochreiter1997long} with a bi-directional extension.
A LSTM model contains self-loops which can keep the gradient flow for long periods. The weights in the self-loops are updated based on the context and can be changed dynamically according to the input sequence by four components of LSTM structures: \\
1) Input gate decides which values will be updated, which depends on the current input $x_{i}$ and the previous hidden state $h_{t-1}$, and is calculated as follows:
\begin{eqnarray}
\label{eq:input}
 s_{t} =\sigma (b_{i}+W_{xi} x_{t}+W_{hi} h_{t-1}+W_{ci}c_{t-1}). \qquad
\end{eqnarray}
2) Forget gate decides what kind of information should be abandoned from the cell state, and is computed as follows:
\begin{eqnarray}
\label{eq:forget}
 f_{t} =\sigma (b_{f}+W_{xf}x_{t}+W_{hf} h_{t-1}+W_{cf}c_{t-1}).  \qquad
\end{eqnarray}
3) Units $c_{t}$ will be updated from the old state $c_{t-1}$ as follows:
\begin{eqnarray}
\label{eq:update}
c_{t} = f_{t}c_{t-1}+i_{t}tanh(W_{xc}x_{t}+W_{hc}h_{t-1}+b_{c}).  \qquad
\end{eqnarray}
4) Output gate is achieved as follows:
\begin{eqnarray}
\label{eq:output1}
o_{t} = \sigma(W_{xo}x_{t}+W_{ho}h_{t-1}+W_{co}c_{t}+b_{o}). \qquad
\end{eqnarray}
\begin{eqnarray}
\label{eq:output2}
h_{t} = o_{t}tanh(c_{t}) \qquad
\end{eqnarray}
where $x_{t}$ represents the current input, $h_{t-1}$ denotes the previous hidden state, W and b are the weight and bias matrices, respectively.

\begin{figure}
\centering
\includegraphics[width=0.48\textwidth]{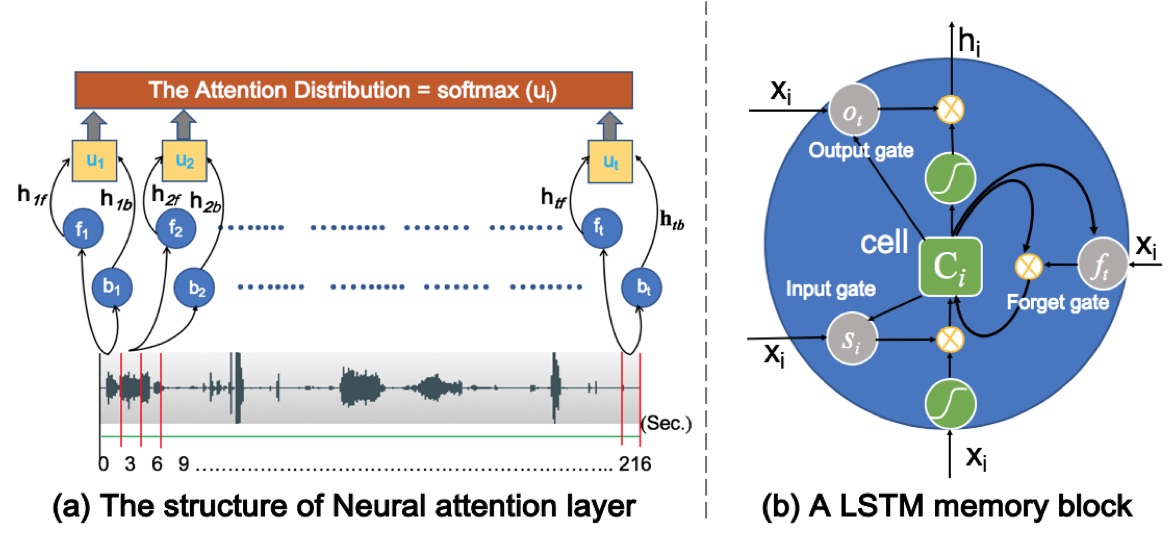}
\caption{\label{models}(a) Main structure of neural attention model, which takes a sequence of audio chunks as input, processes it by the forward and backward LSTM models (achieved by the blue circles), and finally uses the output of bi-directional LSTM models to calculate the attention score and attention distribution as a 72-dimensional vector. (b) A LSTM memory block, including three gates.}
\label{models}
\end{figure}

\begin{figure}
\centering
\includegraphics[width=0.48\textwidth]{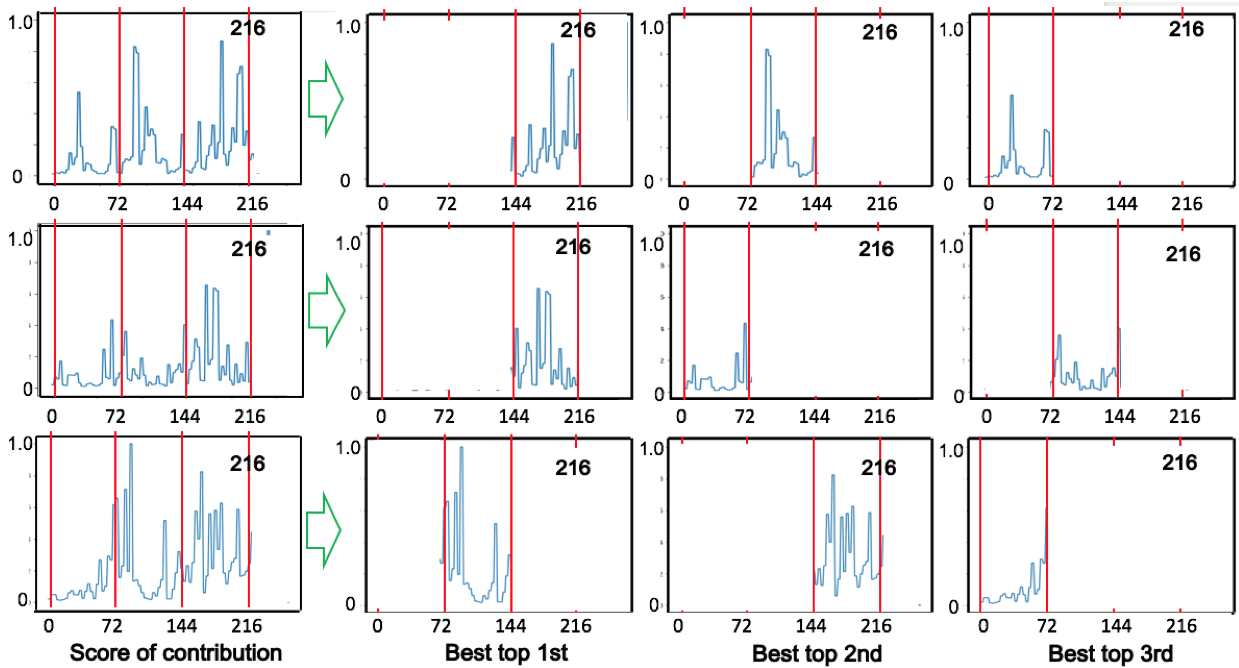}
\caption{\label{scores}Emotion learning model for evaluating the contribution of each chunk to emotions. When an original 216 seconds audio is divided into 3 chunks, the model calculates the contribution score of each chunk, which helps to obtain the top k-th chunk.}
\label{scores}
\end{figure} 

LSTM is a one way computation method. In order to consider both past and future information, the extension of LSTM networks adds one more layer with the opposite temporal sequence and is named bi-directional LSTM, as shown in Fig.~\ref{models}. In our works, each audio is divided into 72 chunks, each with 3 seconds. Then, the bi-directional LSTM model is applied on each chunk. In the attention model, the input of bi-directional LSTMs is the output of global max-pooling layer, which is the first attention layer to compute the contribution scores of different audio chunks. The attention score $u_{t}$ of the t-th chunk can be computed as follows.\\
\begin{eqnarray}
\label{attention}
u_{t} = W^{T} tanh(W_{f}h_{tf} + W_{b}h_{tb}+\beta),
\end{eqnarray}
where $h_{tf}, h_{tb}$ are the outputs of forward and backward LSTM for the t-th chunk, the $W^{T}, W_{f}, W_{b}$ and $\beta$ are the weight parameters of attention score function. When the attention score is obtained, the attention distribution $\theta$ is calculated by a softmax function:

\begin{eqnarray}
\label{attentions}
\theta = softmax(u_{t}).
\end{eqnarray}

We regard this architecture as an emotion learning model~\cite{huang2017music}, which is trained over the MER31K dataset, using emotion tags from AllMusic\footnote{http://www.allmusic.com/moods}. The detail of selecting audio segments achieved by emotion learning model is shown in Fig.\ref{scores}. Firstly, the emotion learning model is used to evaluate the contributions of each chunk to emotions. The contribution score allows us to rank the chunks. Secondly, in the ranked chunks, the best top k are selected. For instance, the first audio in the Fig.~\ref{scores} is divided into 3 chunks, and depending on the contribution scores, the third chunk is selected as the best one, because it has the highest score within the audio.

\begin{figure*}
\centering
\vspace{-2mm}
\includegraphics[width=14.6cm, height=6.7cm]{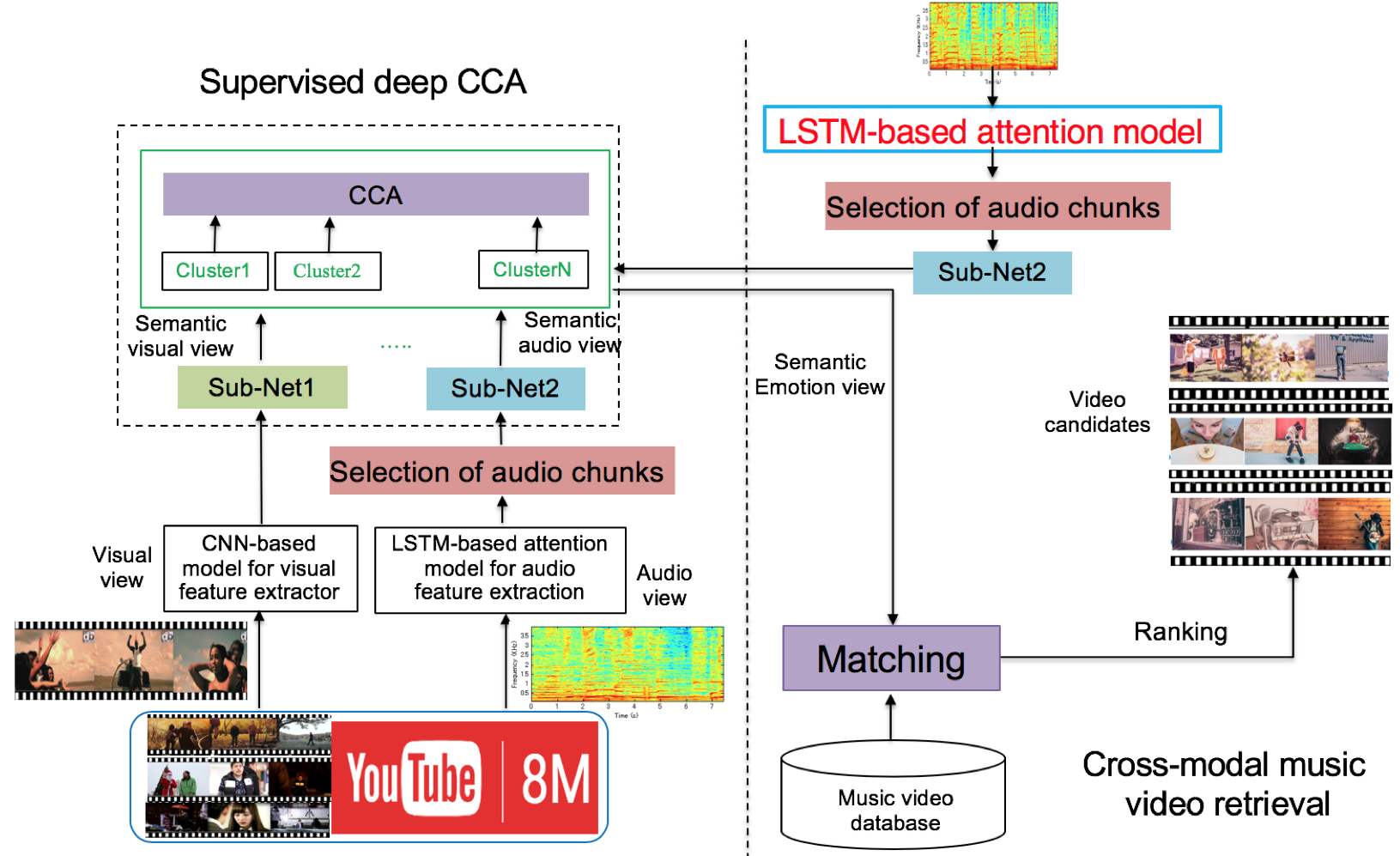}
\caption{\label{figure-2}Audio-visual embedding architecture through S-DCCA. (left) During the training process, the model learns the correlation between audio and visual content. (right) Using audio chunks as input to retrieve music videos.}
\label{architecture}
\vspace{-1mm}
\end{figure*}

\subsection{Supervised Deep Canonical Correlation Analysis and Distance Similarity}
CCA\cite{andrew2013deep} is a classical approach for correlation analysis among two or more modalities. Its core idea is to learn projection matrices that map features of different modalities into the same space, where the correlation between similar items of different modalities are maximized.

Denote $X\in R^{k}$ as an audio feature, $Y\in R^{l}$ as a visual feature, and
denote $W_{x}$, and $W_{y}$ as matrices that linearly map $X$ and $Y$ to the same space, then $W_x$ and $W_y$ are found by maximizing the
correlation between $W_{x}^{T}X$  and $W_{y}^{T}Y$ , as follows:

\begin{eqnarray}
(W_{x}, W_{y}) = \arg\max_{(W_{x}, W_{y})} \frac{W^{T}_{x}\Sigma_{xy}W_{y}}{\sqrt{W^{T}_{x}\Sigma_{xx}W_{x}\cdot W^{T}_{y}\Sigma_{yy}W_{y}}} 
\end{eqnarray}
where $\Sigma_{xx}$ and $\Sigma_{yy}$ represent the covariance matrices of X and Y, respectively and $\Sigma_{xy}$ is their cross covariance matrix.

DCCA extends CCA, realizing non-linear projections by deep
neural networks (DNN). Assume the output of $(i-1)^{th}$ layer is $X_{i-1}$ and $Y_{i-1}$ ($X_{0}=X$ and $Y_{0}=Y$),  and $W_{xi}, W_{yi},  b_{xi}, b_{yi}$ are the weights and biases of the $i^{th}$ layers. Then, the $i^{th}$ layer outputs $X_{i}$ = $s(W_{xi}^T X_{i-1} + b_{xi})$, $Y_{i} = s(W_{yi}^T Y_{i-1} + b_{yi})$ at two branches, where $s$: $R$ $\rightarrow$ $R$ is a nonlinear function. 
The output of the final ($d^{th}$) layer are  $f_{x} = s(W_{xd}X_{d-1} + b_{xd})$, $f_{y}$ = $s(W_{yd}Y_{d-1} + b_{yd})$. Let $\theta_{x}$ represent the parameters $W_{xi}$,  $b_{xi}$, $i$ = 1, ..., d, and $\theta_{y}$ represent the parameters $W_{yi}$, $b_{yi}$, $i$ = 1, ..., d. They are optimized by

\begin{eqnarray}
\label{corr}
(\theta_{x}^{*}, \theta_{y}^{*}) = \arg\max_{(\theta_{x}, \theta_{y})}corr(f_{x}(X,\theta_{x}), f_{y}(Y,\theta_{y})).
\end{eqnarray}
Supervised deep CCA does not merely consider one-to-one
match between all pairs of audio-visual data and apply deep CCA to learn the correlation. In order to preserve the similarity among items with the same class label, audio and visual contents from different videos with the same class label are formed as new relevant pairs to increase the number of training samples. 

In the training process,  maximizing the CCA objective function $G(W_{x}^{T}\Sigma_{xy} W_{y})$ to obtained the linear projections weight $W_{x}$, $W_{y}$ and non-linear function $f_{x}$, $f_{y}$ as follow.

\begin{equation}
\label{corr}
\begin{split}
(W_{x}, W_{y}, f_{x}, f_{y}) = \operatorname*{arg\,max}_{(W_{x}, W_{y}, f_{x}, f_{y})}G(W_{x}^{T}\Sigma_{xy} W_{y}),\\
s.t. W_{x}^{T}\Sigma_{xx} W_{x} = I, W_{y}^{T}\Sigma_{yy} W_{y} = I.
\end{split}
\end{equation}\\
where the covariance matrices $\Sigma_{xx}$, $\Sigma_{xy}$ and $\Sigma_{yy}$ are computed as.





where $N$ is the number of all pairs. The $\sigma$ value decide two factor of the number of training dataset, different from DCCA, S-DCCA considers pairs between audio and visual contents from videos with the same class label, including those pairs formed from different videos, as shown in (\ref{corr3b}). similar to DCCA, all parameters are optimized by formulation (\ref{corr3a}). The left side of  Fig.~\ref{architecture} shows the whole process.

\subsection{K-means clustering}
k-means clustering is a very popular unsupervised learning method for cluster analysis in data mining. k-means clustering enables n variables to be separated into k clusters based on the nearest mean, where k is usually pre-defined by users. 

Given a set of variables X=($x_{1}$, $x_{2}$, …, $x_{n}$), where each variable $x_{i}\in X$ is a d-dimensional vector. In order to cluster them into k groups $G={g_{1}, g_{2}, ..., g_{k}}$ ($k<n$) , firstly, a common method is to randomly choose k values from $X$ as initial cluster centers, then iteratively update the cluster center after assigning each variable $x_{i}$ to its closest cluster till the cluster center never changes. The objective function is defined as follows:
\begin{eqnarray}
\label{kmeans}
\centering
\arg\max_{G}\sum_{i=1}^{k}\sum_{x\in g_{i}}||x-u_{i}||^{2} 
\end{eqnarray} \\
where $u_{i}$ is the mean of points or cluster center of $G_{i}$. In our experiments, we allocate 3 annotated audios for each 10 predefined categories (angry, tender, bitter, cheerful, fun, bright, happy, anxious, calm and warm) to compute the initiated mean $u_{0}$. We use the k-means method to cluster all audios into 10 semantic classes based on the emotion features.
\subsection{Matching and Ranking}
It is not easy to recognize emotion inside the visual modality, because the visual feature of the dataset is high-level semantic features without clear emotion expression like facial expression changes or body movement. However, the high-level semantic information extracted or trained from complicated deep network is able to represent emotion attributes contained in music. Based on this background, we design a S-DCCA model to learn the correlation between audio and video, which enables us to use audio to retrieve video clip. 

The audio-visual embedding is to map audio chunks and visual features to a common space. This space links audio chunks and visual feature in terms of emotion, and enables us to implement cross-modal music video retrieval based on emotion similarity. In the cross-modal  retrieval, given an audio chunk or multiple chunks as query, we calculate the similarity between the query audio chunks and each of the visual features from the database in the emotion-based embedding space. We use the cosine similarity between $f_{x}(X, \theta_{x})$ and $f_{y}(Y, \theta_{y})$ as the similarity metric, which is defined as follows.

\begin{eqnarray}
\label{cos}
Cos(f_{x},f_{y}) = \frac{f_{x}f_{y}}{||f_{x}||.||f_{y}||}
\end{eqnarray}

The detail of our architecture is shown in Fig.~\ref{architecture}. which consists of 2 branches: audio branch and visual branch. Firstly, the pre-trained VGG16 model is used to extract frame-level audio feature and the pre-trained Inception model is used to extract frame-level visual feature, for all data in the dataset. Secondly, the frame-level visual feature is represented as video-level feature by the max pooling method. As for audio branch, we load frame-level audio feature into the pre-trained emotion learning model~\cite{huang2017music} to extract emotion features , based on which the best top k chunks are selected to do music video retrieval, then feed them into Sub-Net1 and Sub-Net2 respectively. Thirdly, based on the extracted emotion features, we apply k-means to cluster the audio into 10 groups. Fourthly, the visual video-level feature and emotion of top k audio chunks are fed into 4 fully connected layers, which generates compact features. Finally, CCA components of these compact features are used to compute the similarity between video and audio chunks.

\section{Experiments}
The performances of the proposed S-DCCA for cross-modal music video retrieval are evaluated in this section, with the studies on the influence of the number of chunks and cross-modal music video retrieval by audio.

\subsection{Dataset and Evaluation Metric}

\subsubsection{\textbf{Dataset}}
The second version of YouTube-8M dataset ~\cite{abu2016youtube} is a large scale video dataset, which includes more than 7 million videos with 4716 classes labeled by the annotation system. The dataset consists of three parts: training set, validate set, and test set. In the training set, each class contains at least 100 training videos. Features of these videos are extracted by the state-of-the-art popular pre-trained models and released for public use. Each video contains audio and visual modality. Based on the visual information, videos are divided into 24 topics, such as sports, game, arts\&entertainment, etc. Specially, the arts\&entertainment topic contains the “music video” label which allows us to construct a music dataset. A video that is included in our music video dataset (MV-10K) should satisfy two conditions:
\begin{enumerate}
\item Each video should include the [music video] label, without other labels.
\item The length of each video ranges from 213 to 219 seconds.
\end{enumerate}
In order to keep enough information in each chunk, the number of chunks for each audio is set as 3, 6, 9. We select videos whose length is around 216 second, because 216 is the common multiple of 3, 6, 9. In our experiment, we separately get 4 subsets of videos based on different video lengths, and the details are shown in Table ~\ref{tab:dataset}.

YouTube-8M has already released the frame-level feature and
video-level feature for both audio and visual information. Frame-level visual feature is extracted by public Inception model which is trained on the ImageNet. Each frame of the visual content is computed per second in the first 6 minutes. After transfer learning and feature dimension reductions with PCA, the dimension of frame-level visual feature is $len$ ×1024, where $len$ is the video lengths in seconds. The video-level visual feature is obtained by the DBoF approach ~\cite{abu2016youtube}.
The frame-level audio feature is extracted by a VGG-like model, as described in ~\cite{hershey2017cnn}, and their average is computed as the video-level audio feature.

\subsubsection{\textbf{Evaluation Metrics}}
In this paper, we choose recall, precision, and MAP as the main metrics for the quantitative evaluation of our method.

\textbf{Precision and Recall}~\cite{powers2011evaluation} are a pair of metrics, which are related to the numbers of relevant documents and retrieved documents. In our experiments, precision is the fraction of retrieved music videos that are relevant to the audio query and recall is the fraction of the relevant music videos that are correctly retrieved.

\textbf{Mean Average Precision (MAP)}~\cite{feng2014cross} for all audio queries is the mean of the average precision for each audio query. When using a music audio as query, in its $N$ ranked retrieved music videos, the average precision (AP) is defined as
\begin{eqnarray}
\label{map}
AP = \frac{1}{R}\sum_{i=1}^{N} p(i)\cdot rel(i)
\end{eqnarray} \\
where $R$ is the number of relevant music videos that belong to the same cluster as the query, $p(i)$ is the precision of top $i$ music videos, $rel(i)$ is a binary value which is 1,  if the $i^{th}$ music video belongs to the same cluster as the query, and 0 otherwise. The cluster for each audio-visual pair only is used in the process of training. During testing, we assume all the music videos that have the same cluster label as the query audio are relevant.
\begin{table}
\centering
  \caption{The Information of Music Dataset Selected}
  \label{tab:dataset}
  \setlength{\tabcolsep}{8mm}{
  \begin{tabular}{cl}
    \toprule
    Length Span&Selected size\\
    \midrule
    216$\pm$3: [213, 219] &  10,000 \\
    216$\pm$6: [210, 222]&  20,000 \\
    216$\pm$9: [207, 225] &  30,000\\
    216$\pm$12: [204, 228] &  40,000\\
    \bottomrule
  \end{tabular}}
\end{table}
\subsection{Experiment Setting}
The frame-level video feature in YouTube-8M is computed one frame per second, according to the pre-trained emotion learning model. We divide the 216 second frame-level audio feature into 72 chunks.The attention model is applied to each chunk to calculate the contribution score of emotion, and each 3 second share the same score. Finally, the result of max pooling is regarded as the score of emotion for each chunk.

The following parameters are used in our experiments:
\begin{itemize}
  \item Network parameter. Both the audio and the branch have 4 hidden layers. The number of units per layer is 512, 512, 256, 256 in the visual branch, and 128, 128, 64, 64 in the audio branch. The number of CCA component is 30. We set the probability of dropout to 0.2 and apply $tanh$ as the
activation function in each hidden layer and use $sigmoid$ function in the final layer.
  \item Experiment parameter. Train batch size is 512 and test batch size is 64. 
The number of training epochs is 50.
  \item We run the experiments with 5 fold cross-validation and get the average performance.
  \item The $RMSProp$ optimizer is used and the learning rate is set to 0.001.
\end{itemize}

\subsection{Baseline}
\textbf{Multi-view}~\cite{zhao2017multi} learning is a technology in machine learning that learn one function per view to model multiple views and optimizes all functions to remove the cross-view gap.

\textbf{CCA}~\cite{thompson2005canonical} algorithm is to find the correlations between two multivariate sets of vectors by linear projections, which depends on singular value decomposition.

\textbf{KCCA}~\cite{cristianini2000introduction} is also a method to extract common features from two data sets Instead of the linear correlation KCCA tries to obtain non-linear correlation through the kernel method, which uses Gaussian kernel and set parameter $\beta$=0.4.

\textbf{DCCA}~\cite{andrew2013deep} is to learn the nonlinear transformations of two data sets such that outputs are highly correlated.

\textbf{C-CCA}~\cite{rasiwasia2014cluster} (Cluster-CCA) is a CCA variant. Different from standard CCA. C-CCA algorithm clusters each data set into several groups or classes and tries to enhance the intra-cluster correlation.

\begin{figure}
\centering
\includegraphics[width=0.48\textwidth]{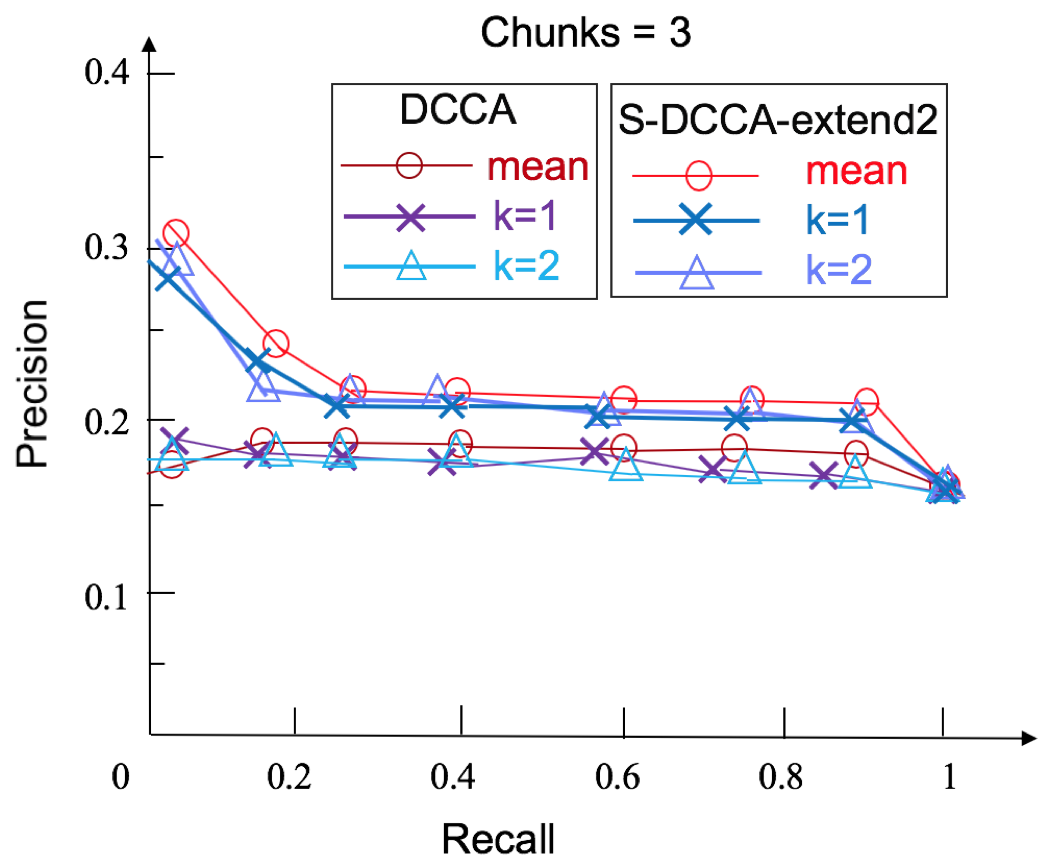}
\caption{\label{dcca}Precision-recall curve with the number of chunks set to 3, where ``mean" denotes using the average of frame level audio feature as query, k (=1, 2) is the number of audio chunks selected as query.}
\label{pr3}
\end{figure}

\begin{figure}
\centering
\includegraphics[width=0.48\textwidth]{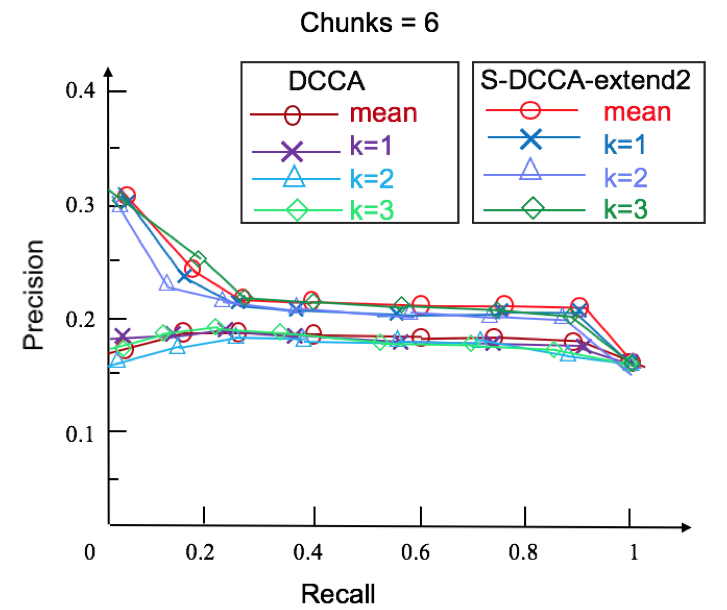}
\caption{\label{pr6}Precision-recall curve with the chunks=6, where  ``mean" denotes using the average of frame level audio feature, k(=1, 2, 3) is the number of audio chunks selected as query.}
\label{pr6}
\end{figure}

\begin{figure}
\centering
\includegraphics[width=0.48\textwidth]{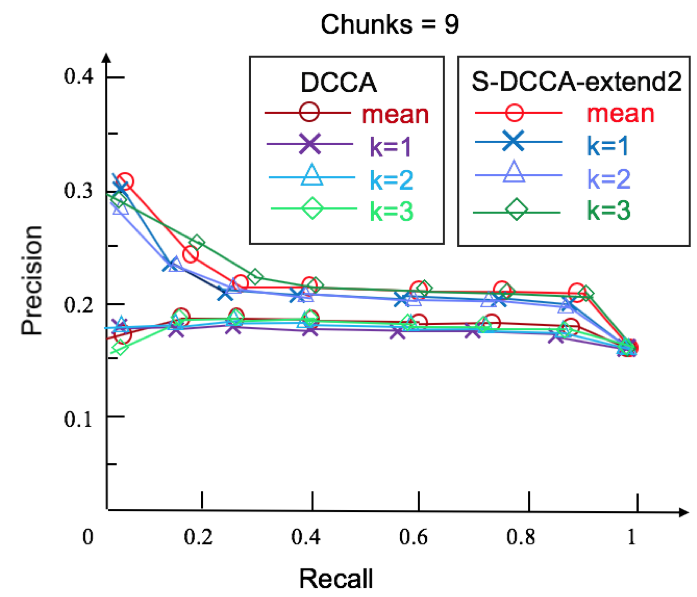}
\caption{\label{dcca}Precision-recall curve with the chunks=9, where  ``mean" denotes using the average of frame level audio feature, k (=1, 2, 3) is the number of audio chunks selected as query.}
\label{pr9}
\end{figure}

\begin{figure}
\centering
\includegraphics[width=0.48\textwidth]{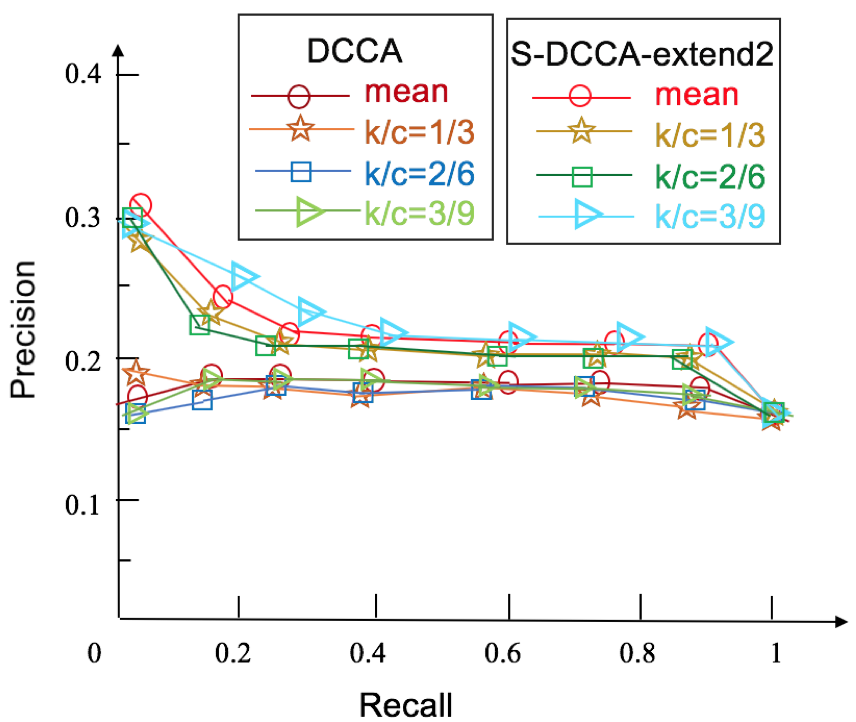}
\caption{\label{dcca} Precision-recall curve, achieved by changing the number of output, where k (=1, 2, 3) is the number of chunks selected from all chunks (c) of an audio as query; for example, k/c=1/3 denotes selecting 1 chunk from an audio that is divided into 3 chunks. "mean" denotes using the average of the whole audio as query.}
\label{prs}
\end{figure}

\begin{figure}
\includegraphics[width=0.48\textwidth]{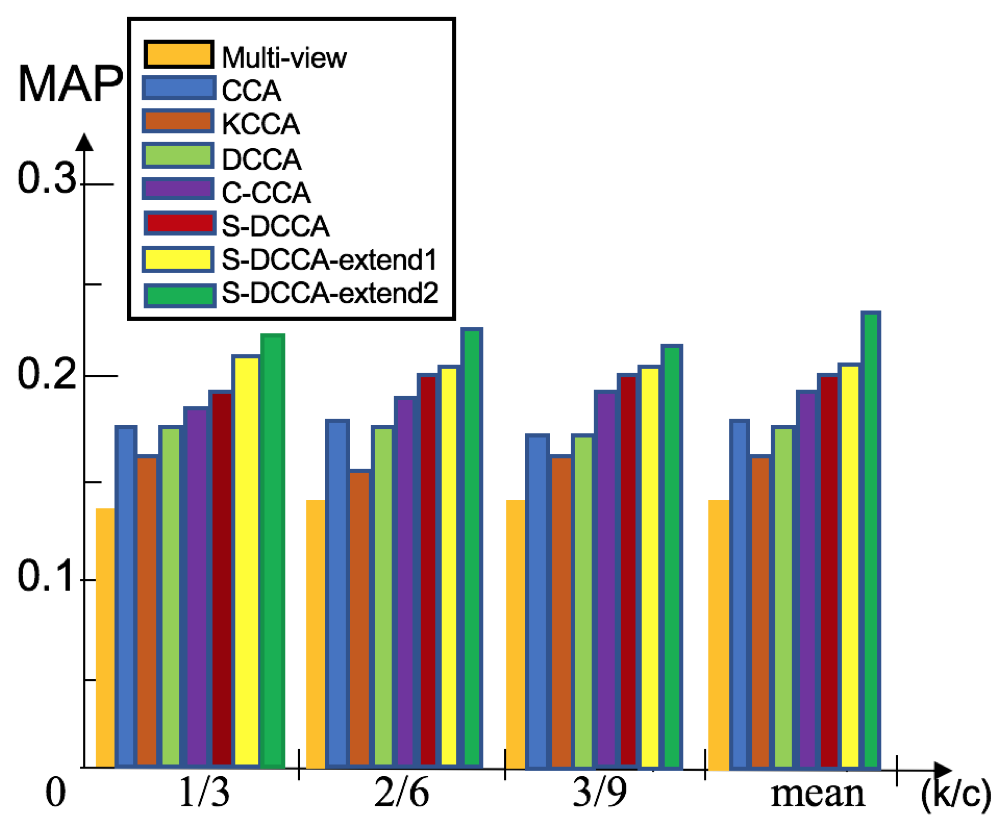}
\caption{\label{map}Mean average precision when using different numbers of audio chunks selected as query for video retrieval, $k$ denotes the number of chunks selected as query, $c$ denotes the number of overall chunks that  the audio is divided into.}
\label{map}
\end{figure}

\begin{table}
  \setlength{\tabcolsep}{4mm}{
  \caption{The MAP results of different methods under different configurations.}
  \label{tab:results}
  \begin{tabular}{ccc cl}
    \toprule
\textbf{k/chunks}& \textbf{1/3} & \textbf{2/6} & \textbf{3/9} & \textbf{mean}
\\ 
\midrule
Multi-views    &14.02   &14.36  &14.25  &14.58 \\
CCA            &18.34   &18.39  &18.32  &18.35 \\
KCCA           &17.54   &17.04  &17.49  &17.80 \\
DCCA           &18.35   &18.39  &18.22  &18.40 \\
C-CCA          &18.51   &19.60  &19.73  &19.72 \\
\textbf{S-DCCA} &\textbf{21.38}  &\textbf{21.43}  &\textbf{21.24}  &\textbf{21.76} \\
  \bottomrule
  \end{tabular}}
\end{table}
\subsection{Experiment Result and Analysis}
Our experiments of S-DCCA use three different training data sets to obtain three different models. The basic C-CCA and S-DCCA model are trained by the 8000 one-to-one pairs. To enhance to intra-cluster correlation, we further consider the correlation between audios and visual contents from different videos of the same cluster, to learn the relationship between the two modalities. We also try to construct more audio-visual pairs during the training. The C-CCA-extend1 and S-DCCA-extend1 are trained by around 0.8 million pairs,  C-CCA-extend2 and S-DCCA-extend2 models by around 1.5 million pairs.
where the former -extend1 model uses 50\% of all music videos of a cluster to form training pairs with each audio in the cluster, and the latter -extend2 model applies 100\% of all music videos in the same cluster to form training pairs.

We use the precision-recall curve to draw the tendency of results as the number of outputs increases so as to compare our S-DCCA model with DCCA model and S-DCCA-extend2 model. Our model tries to leverage the temporal structure inside the query audio, and each query audio is divided into 3, 6, or 9 chunks, from which k chunks are selected as the actual query. In order to investigate the overall performance of our S-DCCA, we use MAP as the metric and compare S-DCCA with others CCA variants (DCCA, C-CCA, KCCA), we set the same dimension of embedding for all methods, and set the same hidden layers structure for DCCA, S-DCCA, S-DCCA-extend1, and S-DCCA-extend2. The correct retrieved video in the rank list which has the same category as query, otherwise it is incorrect video.

Figs.~\ref{pr3},~\ref{pr6},~\ref{pr9} demonstrates the precision-recall curve, comparing DCCA and S-DCCA-extend2 model. The pair of precision and recall value is achieved by changing the number of music videos output. Generally, with the increase of the number of music videos output, the recall increases and the precision decreases. In the S-DCCA-extend2 model, these three figures show that precision starts with the highest value and then sharply decreases before recall arrives at 0.2, then precision almost remains stable as recall increases to 1.0. As is known, the query and the model as two main factors control the curve trend. As for the query factor, when each audio is divided into 3 or 6 chunks, the precision and recall curves of the selected chunks and full-length audio are very close. But when each audio is divided into 9 chunks, and 3 chunks are selected as query, the performance is better than other configurations when the number of output is small. This infers that the 3 chunks have most contribution of emotion and this kind of information is helpful for cross-modal retrieval. As for the model factor, S-DCCA-extend2 is better than DCCA, which indicates that more videos in the output belong to the same cluster as the query in S-DCCA-extend2, than in DCCA.

We also investigate the influence of the number of overall chunks and the number of chunks selected. Fig.~\ref{prs}, shows that with the same volume of audio information as query, when the audio is divided into 9 chunks and 3 chunks are selected as the query the S-DCCA-extend2 model achieves the best performance (precision ranges from 26.6\% to 23.8\%; recall ranges from 0.20 to 0.41). 

In order to further study the influence of the number of overall chunks and the number of chunks selected as query , the MAP results of different models are compared in Table~\ref{tab:results} and Fig.~\ref{map}.
As for the number of chunks selected, generally there is no big difference in MAP when the same model is used.
When the same audio information is used as query, comparing the MAP results among different models, it shows that the training process explicitly exploiting the cluster information generally outperforms the one without cluster information. 
As a result, S-DCCA (and S-DCCA-extend1, S-DCCA-extend2) and C-CCA (and C-CCA-extend1, C-CCA-extend2) can get higher MAP than Multi-views, CCA, KCCA, and DCCA. It indicates that the correlation learning based on both cluster information and instance features is better than those using instance features only.
With the increases in the volume of the training data, from two groups, group 1: C-CCA, C-CCA-extend1, C-CCA-extend2, and group 2: S-DCCA, S-DCCA-extend1, S-DCCA-extend2, the MAP gets higher and higher. It proves that considering all possible pairs within two data sets for each label cluster can get better performance than one-to-one pairs, and it also illustrates the limited training data cannot well learn the correlation between audio and visual feature in this case. Generally, using parts of audio as queries to do retrieval can get close performance as in this case where full-length audio is used as queries.

\section{Conclusions}
We proposed a supervised deep CCA model to learn a semantic space where audio and visual data from music video, which are in different modalities, are linked to learn the cross-modal correlation. Besides the pairwise similarity, the semantic similarity between audio and visual contents from different videos in the same cluster is also explicitly considered. An end-to-end deep architecture that represents an audio sequence as representative chunks is studied. The experimental evaluation run over MV-10K data selected from Youtube-8M proves the effectiveness of the proposed deep audio-visual embedding algorithm in cross-modal music video retrieval. We will try to integrate more users' preference information to our Deep architecture for personalized music cross-modal video recommendation. We will investigate the task of taking a short video as query to retrieve a longer or full audio in the future work.

\section*{ACKNOWLEDGEMENT}
This work was partially supported by JSPS KAKENHI
Grant Number 16K16058. The first Author would like to thank Francisco Raposo for discussing how to implement CCA.
\bibliography{sample-bibliography.bib}
\bibliographystyle{plain}

\end{document}